\documentclass[conference]{IEEEtran}

\usepackage[tight,footnotesize]{subfigure}
\usepackage{float}
\usepackage{url}
\usepackage{latexsym}
\usepackage{epstopdf}
\usepackage[dvipdfm]{graphicx}
\usepackage[usenames,dvipsnames]{color}
\usepackage{amsmath}
\usepackage{amsfonts}
\usepackage{amssymb}
\usepackage{graphicx}
\usepackage{multicol}
\usepackage{verbatim}
\usepackage{enumerate}
\usepackage{cite}
\usepackage{array}
\usepackage{algorithm}
\usepackage{algorithmic}
\usepackage{multirow}

\begin{document}

\title{A Cooperative Q-learning Approach for Real-time Power Allocation in Femtocell Networks}

\author{Hussein Saad\IEEEauthorrefmark{1}, Amr Mohamed\IEEEauthorrefmark{2} and Tamer ElBatt\IEEEauthorrefmark{1}\\
\normalsize \begin{tabular}{cc}
\IEEEauthorrefmark{1}Wireless Intelligence Network Center (WINC), & \hspace{.5in} \IEEEauthorrefmark{2}Computer Science and Engineering Department\\
Nile University, Cairo, Egypt. & \hspace{.5in} Qatar University, P.O. Box 2713, Doha, Qatar.\\
hussein.saad@nileu.edu.eg &  \hspace{.5in} amrm@qu.edu.qa\\
telbatt@nileuniversity.edu.eg \end{tabular}
}

\maketitle

\begin{abstract}

In this paper, we address the problem of distributed interference management of cognitive femtocells that share the same frequency range with macrocells (primary user) using distributed multi-agent Q-learning. We formulate and solve three problems representing three different Q-learning algorithms: namely, centralized, distributed and partially distributed power control using Q-learning (CPC-Q, DPC-Q and PDPC-Q). CPC-Q, although not of practical interest, characterizes the global optimum. Each of DPC-Q and PDPC-Q works in two different learning paradigms: Independent (IL) and Cooperative (CL). The former is considered the simplest form for applying Q-learning in multi-agent scenarios, where all the femtocells learn independently. The latter is the proposed scheme in which femtocells share partial information during the learning process in order to strike a balance between practical relevance and performance. In terms of performance, the simulation results showed that the CL paradigm outperforms the IL paradigm and achieves an aggregate femtocells capacity that is very close to the optimal one. For the practical relevance issue, we evaluate the robustness and scalability of DPC-Q, in real time, by deploying new femtocells in the system during the learning process, where we showed that DPC-Q in the CL paradigm is  scalable to large number of femtocells and more robust to the network dynamics compared to the IL paradigm.
\end{abstract}

%\footnotetext{This work is supported by the Qatar Telecom (Qtel) Grant No.QUEX-Qtel-09/10-10.}

\IEEEpeerreviewmaketitle

\section{Introduction}\label{intro}

Femtocells have been recently proposed as a promising solution to the indoor coverage problem. Although femtocells offer significant benefits to both the operator and the user, several challenges have to be solved to fully reap these benefits. One of the most daunting challenges is their interference on macro-users and other femtocells \cite{4623708}, \cite{femto}. Typically, femtocells are installed by the end user and hence, their number and positions are random and unknown to the network operator a priori. Adding to this the typical dynamics of the wireless environment, a centralized approach to handle the interference problem can not be feasible which, in turn, calls for a distributed
interference management strategies.

Based on these observations, in this paper, we focus on closed access femtocells \cite{closed} working in the same bandwidth with macrocells (i.e. cognitive femtocells), where the femtocells will be the secondary users who try to perform power control to maximize their own performance while maintain the macrocell capacity at certain level. In order to handle the interference generated by the femtocells on the macrocell users, we will use a distributed reinforcement learning \cite{introtoRL} technique called multi-agent Q-learning \cite{q-learning} and \cite{collaborativeRL}. In our context, a prior model of the environment cannot be achieved due to 1) the unplanned placement of the femtocells, 2) the typical dynamics of the wireless enshrinement. In such context, Q-Learning offers significant advantages to achieve optimal decision policies through realtime learning of the environment\cite{2}.

In the literature, Q-learning has been used several times to perform power allocation in femtocell networks. In \cite{1}, authors used independent learning (IL) Q-learning to perform power allocation in order to control the interference generated by the femtocells on the macrocell user. In \cite{2}, authors introduced a new concept called docitive femtocells where a new femtocell can fasten its learning process by learning the policies acquired by the already deployed femtocells, instead of learning from scratch. The policies are shared by Q-table exchange between the femtocells. However, after the Q-tables are exchanged, all the femtocells take their actions (powers) independently, which may generate an oscillating behavior in the system. In \cite{VTC_paper}, we developed a distributed power allocation algorithm called distributed power control using Q-learning (DPC-Q). In DPC-Q, two different learning paradigms were proposed: independent learning (\textbf{IL}) and cooperative learning (\textbf{CL}). It was shown that both paradigms achieves convergence. Moreover, the CL paradigm outperforms the IL one through achieving higher aggregate femtocells capacity and better fairness (in terms of capacity) among the learning femtocells.

However, in \cite{VTC_paper} we did not evaluate the performance of DPC-Q against the networks dynamics, specially after convergence. Also, we did not have any benchmarking algorithm to compare the performance of DPC-Q to. Thus, the contribution of this paper can be summed up as follows:

% any benchmark algorithm. Also, we did not evaluate its robustness against the wireless network environment. Thus,

\begin{itemize}
 \item we propose two new Q-learning based power allocation algorithms: namely, centralized power control using Q-learning (CPC-Q) and partially distributed power control using Q-learning (PDPC-Q). CPC-Q is used for benchmarking purposes, where a central controller, which has all the information about the system (channel gains of all femtocells, system noise, $\cdots$), is responsible for calculating the optimal powers that the femtocells should use. PDPC-Q, which is an agent based power control algorithm, is proposed as: $1)$ it gives the operator the flexibility to work on a global base (e.g. aggregate femtocell capacity instead of subcarrier based femtocell capacity as in DPC-Q), $2)$ it makes DPC-Q comparable to CPC-Q.

     %However, due to the huge overhead needed, CPC-Q is only feasible when few femtocells are deployed.
     %although it is still multi-agent) instead of working on a subcarrier based as in DPC-Q,
\item we evaluate the robustness and scalability of DPC-Q, in both IL and CL paradigms, against two of the dynamics that typically exist in the wireless environment: namely, the random activity of femtocells (when new femtocells are deployed in the system during the learning process) and the density of the femtocells in the macrocell coverage area (the number of femtocells that are interfering on the macro users).
\item we compare our proposed DPC-Q in both IL and CL paradigms to the idea of docitive femtocells presented in \cite{2}.

\end{itemize}
The rest of this paper is organized as follows. In section \ref{sys}, the system model is described. Section \ref{back} presents a brief background about multi-agent Q-learning. In section \ref{problem}, the proposed Q-learning based power allocation algorithms are presented. The simulation scenario and the results are discussed in section \ref{scenario}. Finally the conclusions are drawn in section \ref{conclusion}. \vspace{-2mm}

\section{System Model}\label{sys}

We consider a wireless network composed of one macro cell (with one single transmit and receive antenna Macro Base Station (MBS)) that coexists with $N_{f}$ femtocells, each with one single transmit and receive antenna Femto Base Station (FBS). The $N_{f}$ femtocells are placed indoors within the macrocell coverage area. Both the MBS and the FBSs' transmit over the same $K$ subcarriers where orthogonal downlink transmission is assumed. $U_m$ and $U_f$ macro and femto users are located randomly inside the macro and femto cells respectively. Femtocells within the same range can share partial information during the learning process to enhance their performance.
%The system model is shown in figure \ref{fig.s}.
%
%\begin{figure}[!t]
%\centering
%\includegraphics[height=0.6\columnwidth,width=1\columnwidth]{system_model.eps}
%\caption{Illustration of the considered system model.}
%\label{fig.s}
%\vspace{-5mm}
%\end{figure}

$p_{o}^{(k)}$ and $p_{n}^{(k)}$  denote the transmission powers of the MBS and FBS $n$ on subcarrier $k$ respectively. Moreover, the maximum transmission powers for the MBS and any FBS $n$ are $P_{max}^{m}$ and $P_{max}^{f}$ respectively, where $ \sum_{k=1}^{K} p_{o}^{(k)} \leq P_{max}^{m} $ and $ \sum_{k=1}^{K} p_{n}^{(k)} \leq P_{max}^{f} $.

The system performance is analyzed in terms of the capacity measured in (bits/sec/Hz). The capacity achieved by the MBS at its associated user on subcarrier $k$ is:
\vspace{-0.8mm}
\begin{equation}\label{eq.1}
C_o^{(k)} = \log_{2}(1 + \frac{h_{oo}^{(k)} p_o^{(k)}}{\sum_{n=1}^{N_{f}} h_{no}^{(k)}p_n^{(k)} + \sigma^{2}})
\end{equation}

\vspace{-1mm} where $h_{oo}^{(k)}$ indicates the channel gain between the transmitting MBS and its associated user on subcarrier $k$; $h_{no}^{(k)}$ indicates the channel gain between FBS $n$ transmitting on subcarrier $k$ and the macro user. Finally $\sigma^2$ indicates the noise power. The capacity achieved by FBS $n$ at its associated user on subcarrier $k$ is:
\vspace{-0.8mm}
\begin{equation}\label{eq.2}
C_{n}^{(k)} = \log_{2}(1 + \frac{h_{nn}^{(k)} p_{n}^{(k)}}{\sum_{n'=1,n'\neq n}^{N_{f}} h_{n'n}^{(k)}p_{n'}^{(k)} + h_{on}^{(k)}p_o^{(k)} + \sigma^{2}})
\end{equation}

\vspace{-1mm}where $h_{nn}^{(k)}$ indicates the channel gain between FBS $n$ transmitting on subcarrier $k$ and its associated user; $h_{n'n}^{(k)}$ indicates the channel gain between FBS $n'$ transmitting on subcarrier $k$ and the femto user associated to FBS $n$.

%The aggregate femtocell capacity can then be defined as:
%\vspace{-0.8mm}
%\begin{equation}\label{eq.3}
%C_{femto} = \sum_{n=1}^{N_{f}} \sum_{k=1}^{K} C_{n}^{(k)}
%\end{equation}
\vspace{-2mm}

\section{Multi-agent Q-learning}\label{back}

The scenario of distributed cognitive femtocells can be mathematically formulated using stochastic games \cite{stochasticQ}, where the learning process of each femtocell is described by a task defined by the quintuple $\lbrace N, S, A, P, R(s,\vec{a}) \rbrace$, where:

\begin{itemize}
\item $N = \lbrace 1, 2, \cdots, N_{f} \rbrace$ is the set of agents (i.e. femtocells).
\item $S = \lbrace s_1,s_2,\cdots,s_m \rbrace$ is the set of possible states that each agent can occupy, where $m$ is the number of possible states.
\item $A = \lbrace a_1,a_2,\cdots,a_l \rbrace$ is the set of possible actions that each agent can perform for each task, where $l$ is the number of possible actions.
\item $P$ is the probabilistic transition function that defines the probability that an agent transits from one state to another, given the joint action performed by all agents.
\item $R(s,\vec{a})$ is the reward function that determines the reward fed back to an agent $n$ by the environment when the joint action $\vec{a}$ is performed in state $s \in S$.
\end{itemize}

In the distributed cognitive femtocells scenario, $P$ can not be deduced due to the dynamics of the wireless environment. Thus, one of the most famous techniques that calculates optimal policies without any prior model of the environment is Q-learning . Q-learning assigns each task of each agent a Q-table whose entries are known as Q-values $Q(s_{m},a_{l})$, for each state $s_{m} \in S$ and action $a_{l} \in A$. Thus, the dimension of this table is $m \times l$. The Q-value $Q(s_{m},a_{l})$ is defined to be the expected discounted reward over an infinite time when action $a_{l}$ is performed in state $s_{m}$, and an optimal policy is followed thereafter \cite{q-learning}. The learning process of each agent $n$ at time $t$ can be described as follows: $1)$ the agent senses the environment and observes its current state $s_{m}^{n} \in S$, $2)$ based on $s_{m}^{n}$, the agent selects its action $a_{l}^{n}$ randomly with probability $\epsilon$ or according to: $a_{l}^{n} = arg\max_{a\in A} Q_{n}^{t}(s_{m}^{n},a)$ with probability $1-\epsilon$, where $Q_{n}^{t}(s_{m},a)$ is the row of the Q-table of agent $n$ that corresponds to state $s_{m}^{n}$ at time $t$, and $\epsilon$ is an exploration parameter (a random number) that guarantees that all the state-action pairs of the Q-table is visited at least once, $3)$ the environment makes a transition to a new state $s_{m'}^{n} \in S$ and the agent receives a reward $r_{n}^{t} = R(s_{m}^{n},\vec{a})$ due to this transition, $4)$ the Q-value is updated using equation \ref{eq.4} and the process is repeated.
\vspace{-0.8mm}
\begin{equation}\label{eq.4}
    \begin{split}
    Q_{n}^{t+1}(s_{m}^{n},a_{l}^{n}) := & (1-\alpha)Q_{n}^{t}(s_{m}^{n},a_{l}^{n}) + \\
    & \alpha(r_{n}^{t} +\gamma \max_{a^{'} \in A}Q_{n}^{t}(s_{m'}^{n},a^{'}))
    \end{split}
\end{equation}

\vspace{-1mm} where $\alpha$ is called the learning rate and $0\leq \gamma \leq 1$ is the discount factor that determines how much effect the future rewards have on the decisions at each moment. It should be noticed that the reward $r_{n}^{t}$ depends on the joint action $\vec{a}$ of all agents not on the individual action $a_{l}$. This is the main difference between the multi-agent scenario described here and the single-agent one (when $N_{f} = 1$). In the single-agent case, one of the conditions needed to guarantee that the Q-values converges to the optimal ones is that: the reward of the agent must be dependent only on its individual actions (i.e. the reward function is stationary for each state-action pair) \cite{q-learning},\cite{qprove}. However, for the multi-agent scenario, the reward function is not stationary from the agent point of view, since it now depends on the actions of other agents. Thus, the convergence proof used for the single-agent case can not be used in the multi-agent one. \vspace{-2mm}

%For more details about single-agent and multi-agent Q-learning, the reader is referred to \cite{q-learning}, \cite{collaborativeRL} and \cite{VTC_paper}.
\section{Power Allocation using Q-learning}\label{problem}

In this section, the three proposed Q-learning based power allocation algorithms will be presented:

\subsection{Distributed Power Control Using Q-learning (DPC-Q)}

DPC-Q is a distributed algorithm where multiple agents (i.e: femtocells) aim at learning a sub-optimal decision policy (i.e: power allocation) by repeatedly interacting with the environment. The DPC-Q algorithm is proposed in two different learning paradigms:

\begin{itemize}
\item \textbf{Independent learning (IL)}: In this paradigm, each agent learns independently from other agents (i.e: ignores other agents' actions and considers other agents as part of the environment). Although, this may lead to oscillations and convergence problems, the IL paradigm showed good results in many applications \cite{1}.

\item \textbf{Cooperative learning (CL)}: In this paradigm, each agent shares a portion of its Q-table with all other cooperating agents\footnotemark, aiming at enhancing the femtocells' performance. CL is performed as follows: each agent shares the row of its Q-table that corresponds to its current state with all other cooperating agents (i.e. femtocells in the same range). Then, each agent $n$ selects its action $a_{l}^{n}$ according to the following equation:
    \vspace{-0.8mm}
    \begin{equation}\label{eq.5}
        a_{l}^{n} = arg\max_{a \in A}(\sum_{1 \leq n' \leq N_{f}}Q_{n'}(s^{n'},a))
    \end{equation}

    \footnotetext{We assume that the shared row of the Q-table is put in the control bits of the packets transmitted between the femtocells. The details of the exact protocol lie out of the scope of this paper.}

    \vspace{-1mm}The main idea behind this strategy is explained in details in \cite{VTC_paper}. In terms of overhead, if the number of femtocells is $N_{f}$, then the total overhead needed is $N_{f}.(N_{f}-1)$ messages (each of size $l$) per unit time (i.e. the overhead is quadratic in the number of cooperating femtocells).
    %depends on what is called global Q-value $Q(\textbf{s},\textbf{a})$. This global Q-value represents the Q-value of the whole system (i.e. if the multi-agent scenario is transformed into a single agent one using a centralized controller with global state $\textbf{s}$ and global joint action $\textbf{a}$). This global Q-value can be decomposed into a linear combination of local agent-dependent Q-values: $Q(\textbf{s},\textbf{a}) = \sum_{1 \leq n \leq N_{f}} Q_{n}(s^{n},a^{n})$ \cite{learningAndCooperating}. Thus, choosing the action based on equation \ref{eq.5} would maximize the global Q-value. However, the solution is still not globally optimum because based on equation \ref{eq.5}, all the cooperating agents will choose the same action \cite{VTC_paper}.

\end{itemize}

DPC-Q is an agent and subcarrier based algorithm (i.e. the capacities, states, actions, reward functions are defined for each agent over each subcarrier) \cite{VTC_paper}:

\begin{itemize}
\item \textbf{Agents: }$FBS_n , \forall 1\leq n \leq N_{f}$

\item \textbf{States: } At time $t$ for femtocell $n$ on subcarrier $k$, the state is defined as: $s_{t}^{n,k} = \lbrace I_{t}^{k},\emph{P}_{t}^{n} \rbrace$ where $I_{t}^{k} \in \lbrace 0,1 \rbrace$ indicates the level of interference measured at the macro-user on subcarrier $k$ at time $t$:
    \vspace{-0.8mm}
    \begin{equation}\label{eq.6}
    I_{t}^{k} = \begin{cases} 1,& C_{o}^{(k)} < \Gamma^{o} \\ 0,& C_{o}^{(k)} \geq \Gamma^{o} \end{cases}
    \end{equation}

    \vspace{-1mm} where $\Gamma^{o}$ is the target capacity determining the QoS performance of the macrocell. We assume that the macrocell reports the value of $C_{o}^{(k)}$ to all FBSs through the backhaul connection.

    $\emph{P}_{t}^{n}$ defines the power levels used to quantize the total power FBS $n$ is using for transmission at time $t$:
    \vspace{-0.8mm}
    \begin{equation}\label{eq.7}
    \emph{P}_{t}^{n} = \begin{cases} 0,& \sum_{k=0}^{K} p_{t}^{n,k} < (P_{max}^{f} - A1) \\ 1,& (P_{max}^{f} - A2) \leq \sum_{k=0}^{K} p_{t}^{n,k} \leq P_{max}^{f} \\ 2,& \sum_{k=0}^{K} p_{t}^{n,k} > P_{max}^{f}\end{cases}
    \end{equation}

    \vspace{-1mm} where $A1$ and $A2$ are arbitrary selected thresholds (\emph{several values for $A1$ and $A2$ as well as more power levels were tried through the simulations and the performance gain between these values was marginal}).

\item \textbf{Actions: } The action here is scalar, where the set of actions available for each FBS is defined as the set of possible powers that a FBS can use for transmission on each subcarrier. In the simulations, a range from $-20$ to $P_{max}^{f}$ dBm with step of $2$ dBm is used.

\item \textbf{Reward Functions: } The reward fed back to agent $n$ on subcarrier $k$ at time $t$ is defined as:
    \vspace{-0.8mm}
    \begin{equation}\label{eq.8}
    r_{t}^{n,k} = \begin{cases} e^{-(C_{o}^{(k)}-\Gamma^{o})^2}-e^{-C_{n}^{(k)}} ,&\sum_{k=0}^{K} p_{t}^{n,k}\leq P_{max}^{f} \\ -2,&\sum_{k=0}^{K}p_{t}^{n,k}> P_{max}^{f}\end{cases}
    \end{equation}

    \vspace{-1mm}The rationale behind this reward function is that each femtocell will aim at maximizing its own capacity while: $1)$ maintaining the capacity of the macrocell around the target capacity $\Gamma^{o}$ (convergence is assumed to be within a range of $\pm 1$ bits/sec/Hz from $\Gamma^{o}$), $2)$ not exceeding the allowed $P_{max}^{f}$.

    This reward function was compared to the reward function defined in \cite{VTC_paper}:
    \vspace{-0.8mm}
    \begin{equation}\label{eq.9}
     r_{t}^{n,k} = \begin{cases}e^{-(C_{o}^{(k)}-\Gamma^{o})^2},&\sum_{k=0}^{K}p_{t}^{n,k}\leq P_{max}^{f} \\ -1,&\sum_{k=0}^{K}p_{t}^{n,k}>P_{max}^{f}\end{cases}
    \end{equation}

    \vspace{-1mm} where it was shown that both reward functions maintain the capacity of the macrocell within the convergence range. However, reward function \ref{eq.8} was able to achieve higher aggregate femtocell capacity. In this paper, we show another advantage for reward function \ref{eq.8}, which is: it learns (explores or reacts to network dynamics) better than reward function \ref{eq.9} even when the exploration parameter $\epsilon$ is not used. This mainly depends on the initial value of the Q-values. In this paper, we initially set all the Q-values to zero. Thus, when $\epsilon$ is not used, using reward function \ref{eq.9} will always feed the agent back with a positive reward (given that $P_{max}^{f}$ is not exceeded). Thus, if initially agent $n$ was in state $s_{t}^{n,k}$ on subcarrier $k$ and took action $p_{t}^{n,k}$, the Q-value of this action $Q_{n}(s^{n,k},p^{n,k})$ will be updated using a positive valued reward, thus this Q-value will increase with time, and agent $n$ will keep using the same action forever (since the action is chosen according to the maximum Q-value). Thus, using $\epsilon$ with reward function \ref{eq.9} is a must to have better exploration behavior. On the other hand, using reward function \ref{eq.8} may feed the agent back with positive or negative valued rewards ($e^{-(C_{o}^{(k)}-\Gamma^{o})^2}$ could be smaller than $e^{-C_{n}^{(k)}}$). Thus, given the same initial conditions, agent $n$ could receive a negative valued reward after taking action $p_{t}^{n,k}$, leading to the decrease of its Q-value with time. Once the Q-value decreases below zero, the agent will take another action whose Q-value is greater than the decreased one. Thus, reward function \ref{eq.8} learns (explores) better than reward function \ref{eq.9}.
    %\item The second advantage is that reward function \ref{eq.8} reacts better to the system dynamics than reward function \ref{eq.8}. This advantage will be explained in details in section \ref{results}.
%
%    \end{itemize}

\end{itemize}

In this paper, we also evaluate the robustness and scalability of DPC-Q, in both IL and CL paradigms. We believe that the CL paradigm is much more robust and scalable against the network dynamics compared to the IL paradigm. The reason is that after sharing the row of the Q-table, each femtocell will know the states that all other cooperating femtocells are occupying, and since a state at a certain moment can be defined as: \emph{how the agent sees the environment at that moment}, each femtocell can implicitly know $1)$ how all other femtocells can react to the network dynamics, $2)$ what actions other femtocells are going to take. However, if the femtocells took their actions independently (i.e. IL paradigm), even after knowing the states of each other, oscillating behaviors that may not reach convergence may be generated. One way to overcome this problem is to force the femtocells to make use of the information shared while taking their actions (i.e. taking the actions cooperatively: equation \ref{eq.5}). This could decrease the oscillations in the system, making the femtocells more robust towards the increase of the number of deployed femtocells, and towards the sudden effect caused by any new deployed femtocell.

\subsection{Partially Distributed Power Control Using Q-learning (PDPC-Q)}

PDPC-Q is a partial distributed algorithm, where it is a multi-agent algorithm but only agent dependent (i.e. the states, actions, reward functions are defined for each agent over all subcarriers). As DPC-Q, PDPC-Q works in both IL and CL paradigms. The agents, states, actions and reward functions used for the PDPC-Q algorithm are defined as follows:

%Since, the states in DPC-Q (agent and subcarrier dependent) is not compatible with the definition of states in CPC-Q (global), comparing DPC-Q to CPC-Q will not be feasible. Thus, PDPC-Q is defined as an agent dependent version of DPC-Q in order to be able to compare it to CPC-Q.

\begin{itemize}
\item \textbf{Agents: }$FBS_n , \forall 1\leq n \leq N_{f}$

\item \textbf{States: } At time $t$ the state is defined as: $s_{t} = \lbrace I_{t}\rbrace$ where $I_{t} \in \lbrace 0,1 \rbrace$ indicates the level of interference measured at the macro-user over all subcarriers at time $t$:
    \vspace{-0.8mm}
    \begin{equation}\label{eq.10}
    I_{t} = \begin{cases} 1,& C_{o} < \beta^{o} \\ 0,& C_{o} \geq \beta^{o} \end{cases}
    \end{equation}

    \vspace{-1mm} where $C_{o} = \sum_{k=1}^{K} C_{o}^{(k)}$ is the aggregate macrocell capacity and $\beta^{o}$ is the target aggregate macrocell capacity. 
    
    %Note that since PDPC-Q is not a subcarrier based algorithm, there is no need to define an element in the state to take care of the power used by the FBS as in DPC-Q.

\item \textbf{Actions: } For FBS $n$, the set of actions is defined to be a set of vectors where each vector represents the powers FBS $n$ is using on all subcarriers. 

\item \textbf{Reward Functions: } Reward function \ref{eq.8} can be redefined as:
    \vspace{-0.8mm}
    \begin{equation}\label{eq.12}
    r_{t}^{n} = e^{-(C_{o}-\beta^{o})^2}-e^{-C_{n}}
    \end{equation}

    \vspace{-1mm} where $C_{n} = \sum_{k=1}^{K} C_{n}^{(k)}$ is the aggregate capacity of FBS $n$. Note that since PDPC-Q is not subcarrier based, a power vector in which $P_{max}^{f}$ is exceeded will never be assigned for any FBS. Thus, there is no need to put a negative reward here as in DPC-Q case. The same goes for CPC-Q.
\end{itemize}

\subsection{Centralized Power Control Using Q-learning (CPC-Q)}

CPC-Q is a centralized power control algorithm used to evaluate the performance of our proposed DPC-Q algorithm. CPC-Q can be regarded as the single-agent version of the DPC-Q, and hence, its convergence to the optimal Q-values and thus optimal powers is guaranteed. However, using a centralized controller is not feasible in terms of overhead in multi-agent scenarios. Thus, CPC-Q works only for small scale problems. The agent, states, actions and reward functions used for CPC-Q are defined as follows:

\begin{itemize}
\item \textbf{Agents: } A centralized controller.

\item \textbf{States: } The same as PDPC-Q.

    %Note that there is no need to define a global $\emph{\textbf{P}}_{t}$ in the global state since a centralized controller will never assign powers to any femtocell with $P_{max}^{f}$ is exceeded.

\item \textbf{Actions: } For the central controller, the set of actions is defined to be a set of matrices where each matrix represents the powers of all femtocells over all subcarriers. However, the size of this set grows exponentially with both the number of femtocells and the number of subcarriers. Thus, forming the matrices (all possible actions) from a large set of powers such as the one used in DPC-Q will be infeasible\footnotemark.

     \footnotetext{In the simulations, the set of powers used to form the matrices and the vectors in CPC-Q and PDPC-Q respectively is: $\lbrace 0, 6 ,12 \rbrace$ dBm.}

\item \textbf{Reward Functions: } Since CPC-Q is global, reward function \ref{eq.8} can be redefined as:
    \vspace{-0.8mm}
    \begin{equation}\label{eq.11}
    r_{t} = e^{-(C_{o}-\beta^{o})^2}-e^{-C_{femto}}
    \end{equation}

    \vspace{-1mm} where $C_{femto}$ is defined as $C_{femto} = \sum_{n=1}^{N_{f}} \sum_{k=1}^{K} C_{n}^{(k)}$. 
\end{itemize}

Finally, for the rest of the paper, reward functions \ref{eq.8}, \ref{eq.11} and \ref{eq.12} will be referred to as $R1$, while reward function \ref{eq.9} will be referred to as $R0$. The three proposed algorithms are compared qualitatively in table \ref{table.1}. \vspace{-2mm}

%\footnotetext{Reward functions \ref{eq.8}, \ref{eq.11} and \ref{eq.12} will be referred to as $R1$, while reward function \ref{eq.9} will be referred to as $R0$.}

\begin{table}[!t]
\scriptsize
\caption{Taxonomy of the proposed algorithms.}
\label{table.1}
\centering
\begin{tabular}{|c|l|l|l|l|}
\hline
                & DPC-Q/IL              & DPC-Q/CL              & CPC-Q                 & PDPC-Q            \\
\hline
Complexity      & action is             & action is             & $|A|$  grows          & $|A|$ grows       \\
                & scalar                & scalar                & exponentially         & exponentially     \\
                &                       &                       &in $N_{f}$ and $K$     &in $K$             \\
\hline
Reaction        & Inefficient $\&$      & Efficient $\&$        &-                      & CL is more        \\
to network      & non-robust            & robust                &                       & efficient $\&$    \\
dynamics        &                       &                       &                       & robust than       \\
                &                       &                       &                       & IL                \\
\hline
Scalability     & Inefficient at        & Efficient at         & Infeasible at          & CL is more        \\
                & large $N_{f}$         & large $N_{f}$        & large $N_{f}$          & scalable          \\
                &                       &                      &                        & than IL           \\
\hline
Speed of        & Medium                & Fast                 & Slow                   & CL is faster      \\
Convergence     & convergence           & convergence          & convergence            & than IL           \\
                &                       &                      & since $|A|$            &                   \\
                &                       &                      & is huge                &                   \\
\hline
Overhead        & None                  & $N_{f}^{2}- N_{f}$   &  Huge                  &  CL has larger    \\
                &                       &  messages each       &                        &  overhead than    \\
                &                       &  of size $|A|$       &                        &  IL               \\
\hline
\end{tabular}
\vspace{-6mm}
\end{table}

\section{Performance Evaluation}\label{scenario}

\subsection{Simulation Scenario}

We consider a wireless network consisting of one macrocell serving $U_m = 1$ macro user underlaid with $N_{f}$ femtocells. Each femtocell serves $U_f = 1$ femto-user, which is randomly located in the femtocell coverage area. All of the macro and femto cells share the same frequency band composed of $K$ subcarriers, where orthogonal downlink transmission is assumed. In the simulations, $K$ will change according to the algorithm used: for DPC-Q, $K=6$, while for both CPC-Q and PDPC-Q, $K=3$. The channel gain between any transmitter $i$ and any receiver $j$ on subcarrier $k$ is assumed to be path-loss dominated and is given by:
\vspace{-0.8mm}
\begin{equation}\label{eq.18}
h_{ij}^{(k)} = d_{ij}^{(-PL)}
\end{equation}

\vspace{-1mm} where $d_{ij}$ is the physical distance between transmitter $i$ and receiver $j$, and $PL$ is the path loss exponent. In the simulations $PL = 2$ is used. The distances are calculated according to the following assumptions: $1)$ The maximum distance between the MBS and its associated user is set to $1000$ meters, $2)$ The maximum distance between the MBS and a femto-user is set to $800$ meters, $3)$ The maximum distance between a FBS and its associated user is set to $80$ meters, $4)$The maximum distance between a FBS and another femtocell's user is set to $300$ meters, $5)$ The maximum distance between a FBS and the macro-user is set to $800$ meters.

We used MatLab on a cluster computing facility with $300$ cores to simulate such scenario, where in the simulations we set the noise power $\sigma^2$ to $10^{-7}$, the maximum transmission power of the macrocell $P_{max}^{m}$ to $43$ dBm, the maximum transmission power of each femtocell $P_{max}^{f}$ to $15$ dBm, each of the power levels $A1$ and $A2$ is set to $5$ dBm, the learning rate $\alpha$ to $0.5$, the discounted rate $\gamma$ to $0.9$ and the random number $\epsilon$ to $0.1$ \cite{2} and \cite{VTC_paper}.

\subsection{Numerical Results}\label{results}

\begin{figure}[!t]
\centering
\subfigure[Aggregate femtocells capacity versus the number of femtocells.]{
\includegraphics[height=0.35\columnwidth,width=1\columnwidth]{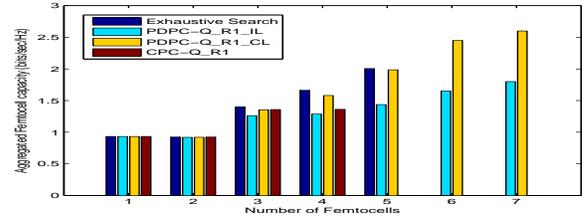}
%\caption{Aggregate femtocells capacity versus the number of femtocells.}
\label{subfig1}
}
%\vspace{-5mm}
\subfigure[Aggregate femtocells capacity versus learning iterations.]{
\includegraphics[height=0.35\columnwidth,width=1\columnwidth]{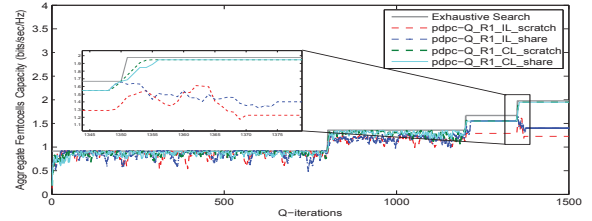}
%\caption{Aggregate femtocells capacity versus learning iterations.}
\label{subfig2}
}
\caption{Aggregate femtocells capacity using CPC-Q and PDPC-Q with $R1$ in both IL and CL paradigms.}
\label{fig.2.4}
\vspace{-6mm}
\end{figure}

Figure  \ref{fig.2.4}\subref{subfig1} shows the aggregate femtocells capacity (as a function of the number of femtocells) using CPC-Q and PDPC-Q with $R1$ in both IL and CL paradigms. It can  be observed that CL is much better than IL, where from the figure it can be shown that the aggregate capacity gain of CPC-Q over PDPC-Q in case of CL is marginal. Since CPC-Q is considered the single agent version of DPC-Q, it should converge to the global optimal values. This is shown in the figure at small number of femtocells ($N_{f} = 1$ and $2$). The optimal values are calculated using exhaustive search over all possible actions, where the optimal value is defined to be the maximum aggregate capacity the system can achieve while maintaining the capacity of the macrocell in the convergence range ($\pm 1$ bits/sec/Hz from $\beta^{o}$). However, starting from $N_{f} = 3$, CPC-Q begins to be infeasible since the size of the possible actions set $A$ becomes very large (at $N_{f} = 5$: $|A| = 320,000,0$). So, besides the computational problems, the condition of visiting all state-action pair becomes infeasible. Thus, getting the optimal value is also not feasible (that's why we stopped CPC-Q at $N_{f} = 4$). Note also that we stopped the exhaustive search at $N_{f} =5$ due to complexity and memory problems, while PDPC-Q is shown at $N_{f} =6$ and $7$ just to illustrate the continuity of our algorithm.

%\begin{figure}[!t]
%\centering
%\includegraphics[height=0.4\columnwidth,width=1\columnwidth]{bar_femto_center.eps}
%\caption{Aggregate femtocells capacity as a function of the number of femtocells using CPC-Q and PDPC-Q with $R1$ in both IL and CL paradigms.}
%\label{fig.1}
%\vspace{-5mm}
%\end{figure}

%Figure \ref{fig.3} shows the fairness among the femtocells (in terms of capacity) using using CPC-Q and PDPC-Q with $RF 1$ in both IL and CL paradigms. Note that the fairness is evaluated using Jain's fairness index \cite{fair}: $f(x_1,x_2,\cdots,x_n) = \frac{(\sum_{i=1}^{n}x_{i})^2}{n\sum_{i=1}^{n}x_i^2}$ where $0 \leq f(x_1,x_2,\cdots,x_n) \leq 1$ and the equality to $1$ occurs when all the femtocells achieve the same capacity. It can be observed that CL achieves better fairness than IL and CPC-Q.
%
%\begin{figure}[!t]
%\centering
%\includegraphics[height=0.5\columnwidth,width=1\columnwidth]{bar_fair_center.eps}
%\caption{Jain's fairness index (in terms of capacity) as a function of the number of femtocells using CPC-Q and PDPC-Q with $RF 1$ in both IL and CL paradigms.}
%\label{fig.3}
%\end{figure}

\begin{figure}[!t]
\centering
\includegraphics[height=0.33\columnwidth,width=1\columnwidth]{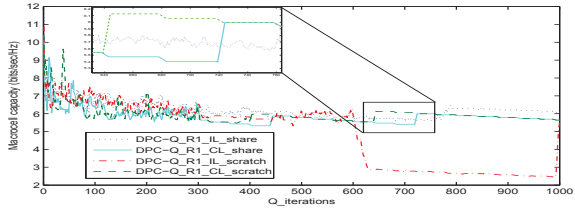}
\caption{Convergence of the macrocell capacity over the Q-iterations on a certain subcarrier where $N_{f}$ was initially $5$, then incremented until $N_{f} = 30$.}
\label{fig.2}
\vspace{-5mm}
\end{figure}

\begin{figure}[!t]
\centering
\includegraphics[height=0.33\columnwidth,width=1\columnwidth]{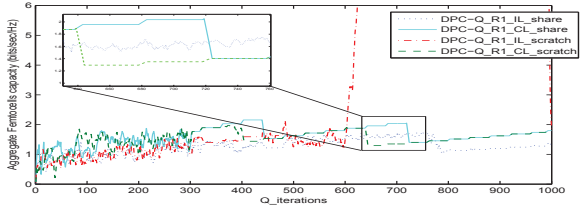}
\caption{Aggregate femtocells capacity over the Q-iterations where $N_{f}$ was initially $5$, then incremented until $N_{f} = 30$.}
\label{fig.3}
\vspace{-5mm}
\end{figure}

Figures \ref{fig.2} and \ref{fig.3} show the robustness of the proposed DPC-Q algorithm. In these figures, we started with $N_{f} = 5$, then we added a new femtocell after every $4000$ iterations\footnotemark to reach $N_{f} = 29$ at the $96000$th iteration. Finally, we add another femtocell at the $99000$th iteration. The figures show how DPC-Q using the CL paradigm is more robust to the deployment of new femtocells compared to the IL paradigm. Moreover, in these figures we compare the performance of DPC-Q to the docitive idea presented in \cite{2}. We investigated two cases: $1)$ the already deployed femtocells share their Q-tables with the new femtocells when they first join the system (suffixed with $\_$share on the figure), $2)$ the new deployed femtocells starts with a zero initialized Q-tables (suffixed with $\_$scratch on the figure). Figure \ref{fig.2} shows the macrocell convergence on a certain subcarrier using DPC-Q with $R1$ in both IL and CL paradigms, where it can be observed that the CL paradigm maintains the macrocell capacity within the range of convergence ($6 \pm 1$ bits/sec/Hz) and reacts well to the effect of the new deployed femtocells, without the need to have a learning phase again every time a new femtocell is deployed, which is a very interesting observation. It can also be observed that our proposed CL paradigm converges to the same value regardless the already deployed femtocells shared its Q-tables with the new ones or not. So, sharing could be ignored, thus decreasing the overall overhead. On the other hand, the IL paradigm showed a very bad reaction to the network dynamics, where $1)$ convergence is not attained (i.e. an oscillating behavior is generated), $2)$ as $N_{f}$ increases, IL paradigm may push the macrocell capacity out of the convergence range when the network becomes more dense. Thus, CL is more scalable than IL. However, it can be noticed that the docitive idea is useful in the IL paradigm, where sharing the Q-tables of the already deployed femtocells with the new ones is much better (in terms of the value that the macrocell capacity oscillates around) than beginning with zero-initialized (scratch) Q-tables. In terms of speed of convergence, it can be noticed that, although the learning process may need large number of iterations initially, CL decreases the dynamics of the learning process, and hence, making it faster. This can be noticed from the figure, where CL converged almost at the $300$th iteration, which is much earlier than the IL paradigm. Also, after the deployment of each new femtocell, CL took less than $10$ iterations only - around $0.01$ seconds - to re-achieve convergence.

\footnotetext{In figures \ref{fig.2} and \ref{fig.3}, $\epsilon$ is removed at the $50000$th iteration and the figures were drawn with step $= 100$ in order to achieve better resolution.}

Figure \ref{fig.3} shows the aggregate femtocells capacity over the learning iterations. It can be noticed that using the CL paradigm, the aggregate capacity increases as more femtocells are deployed in the network, while in the IL paradigms, since convergence is already not maintained, the aggregate capacity behavior has a sporadic behavior, which indicates clearly that IL is not efficient to react to the network dynamics. However, in the CL paradigm from the $64000$th to the $72000$th iteration($640$th to $720$th according to figure's scale), it can be noticed that the aggregate capacity decreases. The reason is that as more femtocells are being deployed, the network becomes very dense and since using the CL paradigm makes the cooperating femtocells use the same powers, this may force the macrocell capacity to violate the range of convergence. Thus, all the femtocells will have to decrease the power used to maintain again the macrocell capacity within the range of convergence leading to the decrease of their aggregate capacity. Note that at the $64000$th and $72000$th iterations, $\epsilon$ is already removed, which proves that $R1$ learns well even when $\epsilon$ is removed.

%Figure \ref{fig.5} sums up the advantages of $R1$ over $R0$, where it shows the macrocell convergence on another subcarrier using DPC-Q with $R1$ and $R0$ in both the IL and CL paradigms. It can be observed that, although using the CL paradigm with $R0$ maintain convergence, it does not react well to the network dynamics and may push the macrocell capacity out of the range of convergence (the blue dotted curve), while with $R1$, CL achieves macrocell capacity within the range of convergence even if it is violated at some time due to the deployment of new femtocells (the green dashed curve). For the other advantage, it can be observed, after $\epsilon$ is removed at the $50000$th iteration, that the new deployed femtocells almost learn nothing and keep transmitting with the same power when $R0$ is used in the IL paradigm (i.e. the macrocell capacity does change with time), while there are some oscillation is the macrocell capacity curve when $R1$ is used, which means that the new femtocells are changing their powers. Thus, $R1$ learns (explores) better than $R0$.
%
%\begin{figure}[!t]
%\centering
%\includegraphics[height=0.5\columnwidth,width=1\columnwidth]{macro2022.eps}
%\caption{Convergence of the macrocell capacity over the Q-iterations on a certain subcarrier where $N_{f}$ was initially $5$, then incremented until $N_{f} = 30$.}
%\label{fig.5}
%\vspace{-5mm}
%\end{figure}

%\begin{figure}[!t]
%\centering
%\includegraphics[height=0.5\columnwidth,width=1\columnwidth]{1st_femto_Power.eps}
%\caption{The power of the first femtocell on a certain subcarrier.}
%\label{fig.6}
%\end{figure}

Finally, in order to compare the aggregate capacity the CL paradigm achieves, after the incremental deployment of femtocells, to the ideal value, we used the small scale problem again. This is shown in figure \ref{fig.2.4}\subref{subfig2}, where we started with $N_{f} = 2$ and added an extra femtocell at the $8000$th, $12000$th and $13500$th iterations\footnotemark. Again, it can be observed that CL achieves aggregate capacity that is very close to the optimal one while the IL paradigm is far from it. \vspace{-2mm}

\footnotetext{In figure \ref{fig.2.4}\subref{subfig2}, $\epsilon$ is removed at the $12000$th iteration and the figure was drawn with step $= 10$ in order to achieve better resolution.}

%\begin{figure}[!t]
%\centering
%\includegraphics[height=0.4\columnwidth,width=1\columnwidth]{femto_center_final3.eps}
%\caption{Aggregate femtocells capacity over the Q-iterations where $N_{f}$ was initially $2$, then incremented until $N_{f} = 5$.}
%\label{fig.4}
%\vspace{-5mm}
%\end{figure}

\section{Conclusion}\label{conclusion}

In this paper, three Q-learning based power allocation algorithms for cognitive femtocells scenario are presented: namely, DPC-Q, CPC-Q and PDPC-Q. Although DPC-Q was presented in previous work, in this paper it is extended, in both of its learning paradigms: IL and CL, to evaluate its performance, robustness and scalability. In terms of performance, DPC-Q is extended to PDPC-Q and then compared to CPC-Q, where the simulations showed that the CL paradigm outperforms the IL and achieves aggregate femtocell capacity that is very close the optimum one. In terms of robustness, the CL paradigm was found to be much more robust against the deployment of new femtocells during the learning process, where the results showed that the CL paradigm outperforms the IL paradigm in: $1)$ maintaining convergence, $2)$ learning better (i.e. reacting better to the network dynamics), especially when a suitable reward function such as the one defined in the simulations is used, $3)$ converging to the target capacity regardless the old femtocells share their experience (i.e. Q-tables) with the new deployed ones or not and $4)$ speeding up the convergence. Finally, in terms of scalability, CL paradigm reacted better to the network dynamics and maintained convergence, even when the number of the femtocells is large . \vspace{-2mm}

\section*{Acknowledgment}

This work is supported by the Qatar Telecom (Qtel) Grant No.QUEX-Qtel-09/10-10.

\bibliographystyle{./IEEEtran}
\bibliography{./bare_conf}

%\end{thebibliography}
\end{document}